# Torsional Waves Propagation in an Initially Stressed Dissipative Cylinder

**M. M. Selim**

Department of Mathematics, Faculty of Education,
Suez Canal University, Suez, Egypt
selim23@yahoo.com

Present address: Department of Mathematics, Al-Aflaj Community College
King Saud University, P.O. Box 710, Al-Aflaj 11912
Saudi Arabia

**Abstract**

The present paper has been framed to show the effect of damping on the propagation of torsional waves in an initially stressed, dissipative, incompressible cylinder of infinite length. A governing equation has been formulated on Biot's incremental deformation theory. The velocities of torsional waves are obtained as complex ones, in which real part gives the phase velocity of propagation and corresponding imaginary part gives the damping. The study reveals that the damping of the medium has strong effect in the propagation of torsional wave. Since every medium has damping so it is more realistic to use the damped wave equation instead of the undamped wave equation. The study also shows that the velocity of propagation of such waves depend on the presence of initial stress. The influences of damping and initial stresses are shown separately.

## 1. Introduction

The studies of wave propagation in an elastic media have received more attention recently, chiefly because the need of a complete understanding of different medium characteristics with respect to mechanical shocks and vibrations is often felt in the Earth. The basic literature on the propagation of elastic waves is the monograph by Ewing et al.[6]. A large number of papers have been published in different journals after the publication of this book.

The problems related to prestressed elastic medium has been a subject of continued interest due to its importance in various fields, such as earthquake engineering, seismology and Geophysics. Significant initial stress way develops in a medium as a result of several physical factors. An early effort by Cauchy [4] used assumption that stress was due to central forces between particles of the solid. Bromwich [3]examined the effects of gravity on surface waves. Southwell [12] discussed the case of uniform initial stress. Love [9] derived the equations for an incompressible solid under hydrostatic pressure. Quite a good amount of information about the theory of initial stresses is contained in the well-known book entitled Mechanics Incremental Deformations by Biot [2].

Although much information is available on the propagation of surface waves, such as, Rayleigh waves, Love waves, and Stonely waves etc., the torsional wave has not drawn much attention and very little literature is available on propagation of this wave. Lord Rayleigh [8], in his remarkable paper, showed that the isotropic homogenous elastic half-space does not allow a Torsional surface wave to propagate. Bhattacharya [1] has been investigated the torsional wave propagation in a two-layered circular cylinder with imperfect bond. The propagation of torsional wave in a finite piezoelectric cylindrical shell has been discussed by Paul and Sarma [10]. Recently, the propagation of torsional wave in an initially stressed cylinder has been discussed by Dey and Dutta [5].

In the above studies the usual torsional wave equation has been used. In many physical applications, one encounters the situation where the usual torsional wave equation does not serve as a realistic model. For instance, the torsional wave equation does not model correctly if the medium offers resistance.

The present paper has been framed to show the effect of damping on the propagation of torsional waves in an initially stressed, dissipative, incompressible cylinder of infinite length. A governing equation has been formulated on Biot's incremental deformation theory. The velocities of torsional waves are obtained as complex ones, in which real part gives the phase velocity of propagation and corresponding imaginary part gives the damping. The study reveals that the damping of the medium has strong effect in the propagation of torsional wave. Since every medium has damping so it is more realistic to use the damped wave equation instead of the undamped wave equation. The study also shows that the velocity of propagation of such waves depend on the presence of initial stress. The dispersion curves for all possible cases have been drawn.

## 2. Theory

Consider a solid cylinder of incompressible elastic material of radius $a$ and of infinite length under initial stress. A cylindrical polar coordinates system $(r,\theta,z)$ is chosen for present study with the z-axis coincide with the axis of the cylinder. The presence of initial stress along z-axis will generate orthotropic symmetry in the medium. The dynamical equation of motion under initial compression stress $S_{zz} = -P$ along the cylinder axis *z* with a damping term are :

$$\frac{\partial s_{rr}}{\partial r} + \frac{1}{r}\frac{\partial s_{r\theta}}{\partial \theta} + \frac{\partial s_{rz}}{\partial z} + \frac{s_{rr} - s_{\theta\theta}}{r} - P\frac{\partial \omega_\theta}{\partial z} = \rho\frac{\partial^2 u}{\partial t^2} + \gamma\frac{\partial u}{\partial t},$$

$$\frac{\partial s_{r\theta}}{\partial r} + \frac{1}{r}\frac{\partial s_{\theta\theta}}{\partial \theta} + \frac{\partial s_{\theta z}}{\partial z} + \frac{2s_{r\theta}}{r} + P\frac{\partial \omega_r}{\partial z} = \rho\frac{\partial^2 v}{\partial t^2} + \gamma\frac{\partial v}{\partial t},$$

$$\frac{\partial s_{rz}}{\partial r} + \frac{1}{r}\frac{\partial s_{\theta z}}{\partial \theta} + \frac{\partial s_{zz}}{\partial z} + \frac{s_{rz}}{r} - P\left[\frac{\partial \omega_\theta}{\partial r} - \frac{\partial \omega_r}{\partial \theta}\right] = \rho\frac{\partial^2 w}{\partial t^2} + \gamma\frac{\partial w}{\partial t}, \tag{1}$$

where $\rho$ is the density, $S_{rr}, S_{\theta\theta}, S_{zz}, S_{rz}, S_{r\theta}$ and $S_{\theta z}$ are the corresponding stress components in their conventional sense, $u, v, w$ are the displacement components in radial, circumferential and axial directions and $\omega_r, \omega_\theta$ are the rational components given by

$$\omega_r = \frac{1}{2}(\frac{\partial w}{\partial \theta} - \frac{\partial v}{\partial z}),$$
$$\omega_\theta = \frac{1}{2}(\frac{\partial u}{\partial z} - \frac{\partial w}{\partial r}) \tag{2}$$

Equation (1) has dissipation or damping term which is proportional to velocity with constant of proportionality being $\gamma$.

In cylindrical coordinates the tress-strain relations for an initially stressed orthotropic elastic medium, are taken as [2].

$$s_{rr} = B_{11} e_{rr} + B_{12} e_{\theta\theta} + B_{13} e_{zz},$$
$$s_{\theta\theta} = B_{21} e_{rr} + B_{22} e_{\theta\theta} + B_{23} e_{zz},$$
$$s_{zz} = B_{13} e_{rr} + B_{23} e_{\theta\theta} + B_{33} e_{zz},$$
$$s_{\theta z} = 2Q_1 e_{\theta z},$$
$$s_{rz} = 2Q_2 e_{rz},$$
$$s_{r\theta} = 2Q_3 e_{r\theta}, \tag{3}$$

where $B_{ij}$ and $Q_i$ $(i, j = 1,2,3)$ are the incremental elastic coefficients and shear modulus, respectively.

The incremental strain components in terms of the displacement components $u, v$ and $w$ are

$$e_{rz} = e_{zr} = \frac{1}{2}(\frac{\partial u}{\partial z} + \frac{\partial w}{\partial r}), \quad e_{rr} = \frac{\partial u}{\partial r}, \quad e_{\theta\theta} = \frac{1}{r}(\frac{\partial v}{\partial \theta} + u), \quad e_{zz} = \frac{\partial w}{\partial z}$$

$$e_{r\theta} = e_{\theta r} = \frac{1}{2}(\frac{1}{r}\frac{\partial u}{\partial \theta} - \frac{v}{r} + \frac{\partial v}{\partial r}), \quad e_{\theta z} = e_{z\theta} = \frac{1}{2}(\frac{\partial v}{\partial z} + \frac{1}{r}\frac{\partial w}{\partial \theta}),$$
$$\tag{4}$$

For the present problem we confine ourselves to the study of torsional wave propagation in damping medium. Following the usual procedure for the problems having $\theta$ symmetry it can easily be seen that the first and the third equations of motion given are automatically satisfied as $u = w = 0$.

Using (2) the remaining equation of motion becomes

$$\frac{\partial s_{r\theta}}{\partial r} + \frac{\partial s_{\theta z}}{\partial z} + \frac{2 s_{r\theta}}{r} - \frac{P}{2}\frac{\partial^2 v}{\partial z^2} = \rho \frac{\partial^2 v}{\partial t^2} + \gamma \frac{\partial v}{\partial t} \tag{5}$$

The stress-strain relationship for this case may be taken as

$$s_{r\theta} = 2 Q_1 e_{r\theta},$$
$$s_{\theta z} = 2 Q_2 e_{\theta z}, \tag{6}$$

where $Q_1$ and $Q_2$ are incremental elastic coefficients and their values have been obtained by Biot [2],

$$Q_1 = \frac{\mu}{2}(\lambda_r^2 + \lambda_\theta^2)$$
$$Q_2 = \frac{\mu}{2}(\lambda_\theta^2 + \lambda_z^2), \tag{7}$$

where $\mu$ is the shear modulus of original unstressed medium and $\lambda_r, \lambda_\theta, \lambda_z$ are extension ratios representing the lengths acquired by the sides of cube originally of the unit dimension oriented along the directions of orthotropic symmetry.

The condition of incompressibility
$$\lambda_r \lambda_\theta \lambda_z = 1 \tag{8}$$

Using the relationship between the initial stresses and extension ratios obtained by Biot[2], one gets

$$\lambda_r^2 = \lambda_\theta^2 = \frac{1}{\lambda_z} = \frac{1}{\lambda} \text{(say)} \tag{9}$$

$$P = \frac{\mu}{\lambda}(1 - \lambda^3) \tag{10}$$

Inserting relations (2), (3), (4), (6), (7), (8), (9) and (10) in (5) one gets

$$\frac{\partial^2 v}{\partial r^2} + \lambda^3 \frac{\partial^2 v}{\partial z^2} + \frac{1}{r}\frac{\partial v}{\partial r} - \frac{v}{r^2} = \frac{\rho \lambda}{\mu}\frac{\partial^2 v}{\partial t^2} + \frac{\gamma \lambda}{\mu}\frac{\partial v}{\partial t} \tag{11}$$

Assuming harmonic wave solution $e^{ik(z-ct)}$, the solution for circumferential displacement in the cylinder becomes

$$v(r,z,t) = V(r) \, e^{ik(z-ct)}. \tag{12}$$

Using eq.(12), eq.(11) takes the form

$$r^2 \frac{d^2V}{dr^2} + r \frac{dV}{dr} + [\eta^2 r^2 - 1]V = 0, \tag{13}$$

where

$$\eta^2 = k^2 (\frac{c^2 \lambda}{\beta^2} - \lambda^3) + \frac{ikc\lambda\gamma}{\mu} \tag{14}$$

and $\beta = (\frac{\mu}{\rho})^{\frac{1}{2}}$ the velocity of shear wave in the initial stress-free medium; $c$ is torsional wave velocity in the initially stressed cylinder.

Equation (13) called the parametric Bessel equation of first order and its general solution is

$$V(r) = AJ_1(\eta r) + BY_1(\eta r), \tag{15}$$

where $J_1(\eta r)$ and $Y_1(\eta r)$ denote Bessel's functions of the first order and of the first and second kind, respectively.

### 3. Boundary Conditions

In the absence of external body forces on the boundary, the surface of the cylinder is kept stress free and also we assume the displacement at $r = 0$ is finite. The boundary conditions may be taken as

$$s_{r\theta} = \mu(\frac{dV}{dr} - \frac{V}{r}) \qquad \text{at} \quad r = a \tag{16}$$

The second part of eq.(15) is left out from the complete solution for well-known physical reasons [11].

Using solution (12) the boundary condition (16) gives

$$\xi J_1'(\xi) - J_1(\xi) = 0 \tag{17}$$

where $\xi = \eta a$ and prime denotes derivatives with respect to r.

Equation(17) gives a multiple roots of $\xi$, the first three being 0, 5.136, 8.418 [5].
The relation (14) gives

$$\frac{c^2}{\beta^2} - \frac{c}{\beta} I - R = 0, \tag{18}$$

where $c$ is torsional wave velocity in the initially stressed cylinder, and

$$I = -i\frac{\delta}{\rho k}, \qquad R = \left[\left(\frac{\xi}{ka}\right)^2 \cdot \frac{1}{\lambda} + \lambda^2\right], \tag{19}$$

where $\delta = \gamma a$ is the damping parameter.

From Eq.(18), the phase velocity ($c$) of torsional wave propagating in the medium may be obtained as

$$\frac{c}{\beta} = \frac{1}{2}\left(I + \sqrt{\Omega}\right), \tag{20}$$

where

$$\Omega = I^2 + 4R. \tag{21}$$

Equation (20) gives the velocity of propagation as well as damping. Real part of the right hand side corresponds to phase velocity and the respective imaginary part corresponds to damping velocity of torsional waves.

The presence of $\delta$ and $\lambda$ in relation (20) shows the effect of damping and initial stresses on the velocity of torsional wave, respectively.

## 4. Particular cases

**4.1 Case1**: For non-dissipative medium ($\delta = 0$), and the velocity of torsional wave is

$$\frac{c}{\beta} = \left[\left(\frac{\xi}{ka}\right)^2 \cdot \frac{1}{\lambda} + \lambda^2\right]^{\frac{1}{2}}, \tag{22}$$

which coincide with the result of Dey and Dutta [5].

**4.2 Case2**: The presence of $\lambda$ in relation (19) shows that the effect of initial stress on the velocity of torsional wave. If the medium is free from initial stresses, $\lambda = 1$ and the velocity of torsional wave is

$$\frac{c}{\beta} = \left[\left(\frac{\xi}{ka}\right)^2 + 1\right]^{\frac{1}{2}}, \tag{23}$$

which coincide with the result given by Kolsky [7].

## 5. Numerical Results and discussion

In order to perform numerical calculation from the frequency equation (20), let us consider a model with $\rho = 2.15$ g/cm$^3$. The numerical values of $c/\beta$ are calculated for different values of damping parameter ($\delta = 0.05$, $0.1$, $0.15$, $0.2$). Also, the values of $c/\beta$ are calculated taking $\xi = 5.136$ and $\xi = 8.418$ respectively, when the medium under initial compression stress ($\lambda < 1$) and without initial stress ($\lambda = 1$). The curves have been plotted for different value of $ka$ and the results of computations are presented in Figs. 1 to 3.

Fig.1 shows the effect of damping on the velocity of torsional wave propagating in a dissipative medium. The curves reveal clearly that the presence of damping affects the propagation of torsional wave in sense that every small change in the value of damping parameter $\delta$ produces a substantial change in the damping velocity.

Fig.2 gives the variation of $c/\beta$ with $ka$ for different values of initial stress parameter $\lambda$ at $\xi = 5.136$. It is clear that from the curves that, the velocity of torsional wave increases as the compressive stress increases.

Fig.3 gives the variation of $c/\beta$ with $ka$ for different values of initial stress parameter $\lambda$ at $\xi = 8.418$. The observation made from Fig.2 is supported by the results shown in this figure.

## 6. Conclusion

It is concluded that the damping of the medium has strong effect in the propagation of torsional waves. Since every medium has damping so it is more realistic to use the damped wave equation instead of the undamped wave equation. The study also shows that the velocity of propagation of such waves depend on the initial stress present in the medium.

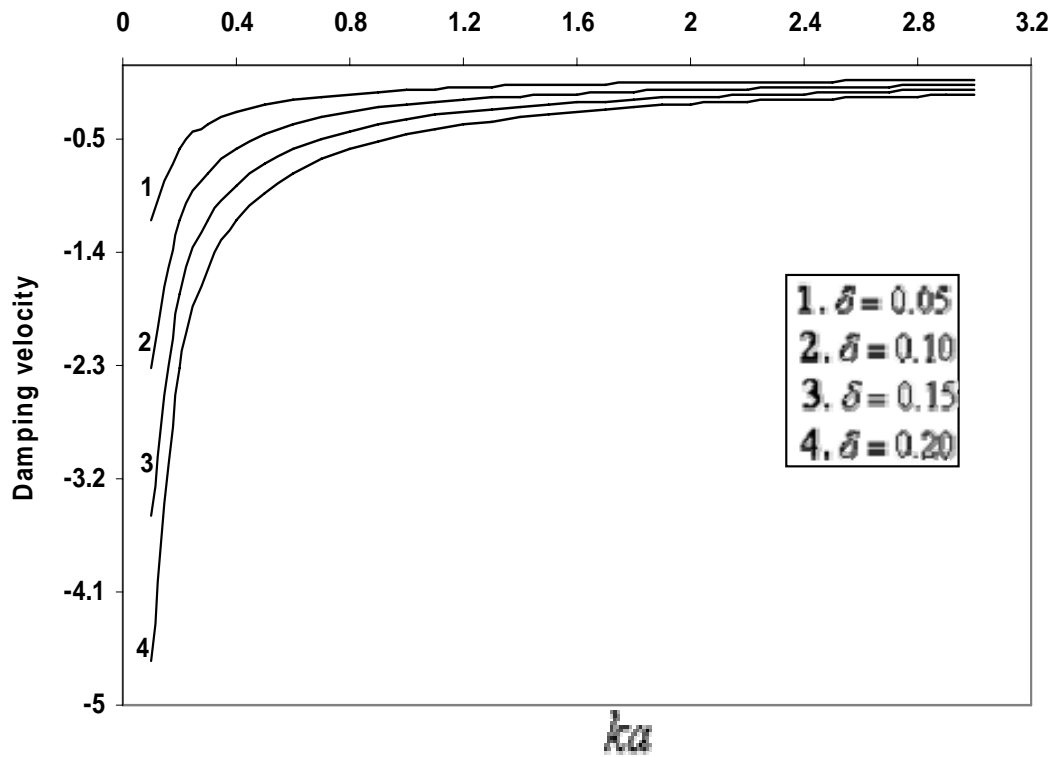

**Fig. 1 Variation of damping velocity versus $ka$ for different values of damping parameter $\delta$.**

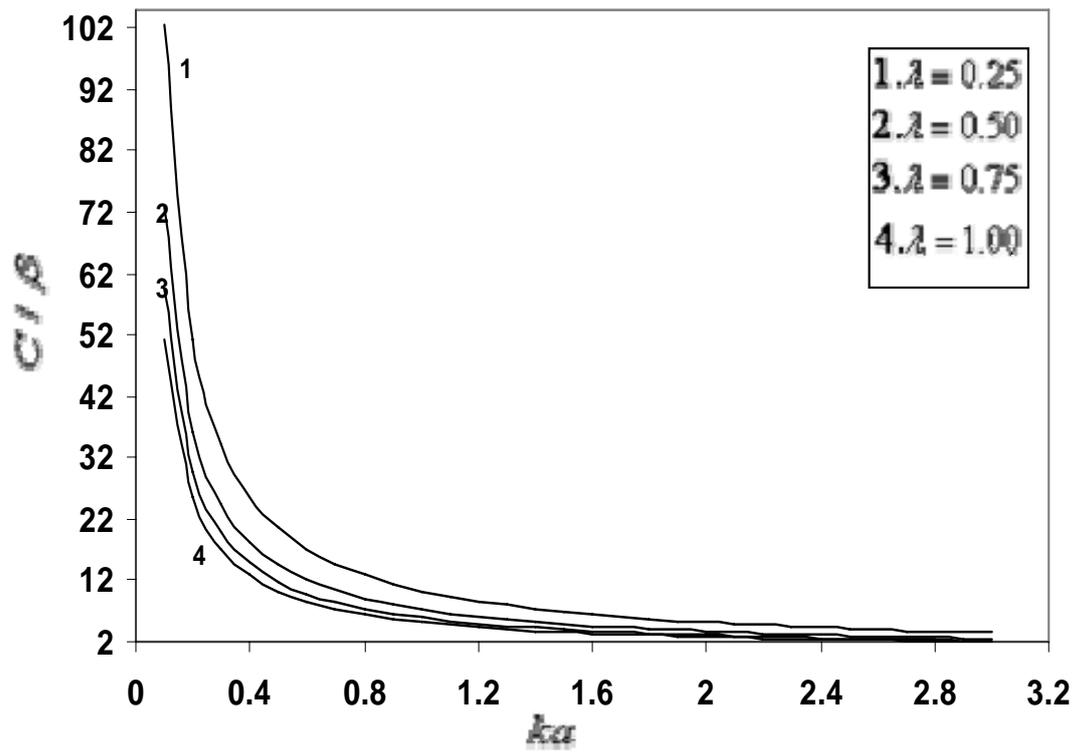

**Fig. 2 Variation of** $c/\beta$ **versus** $ka$ **at** $\xi = 5.136$

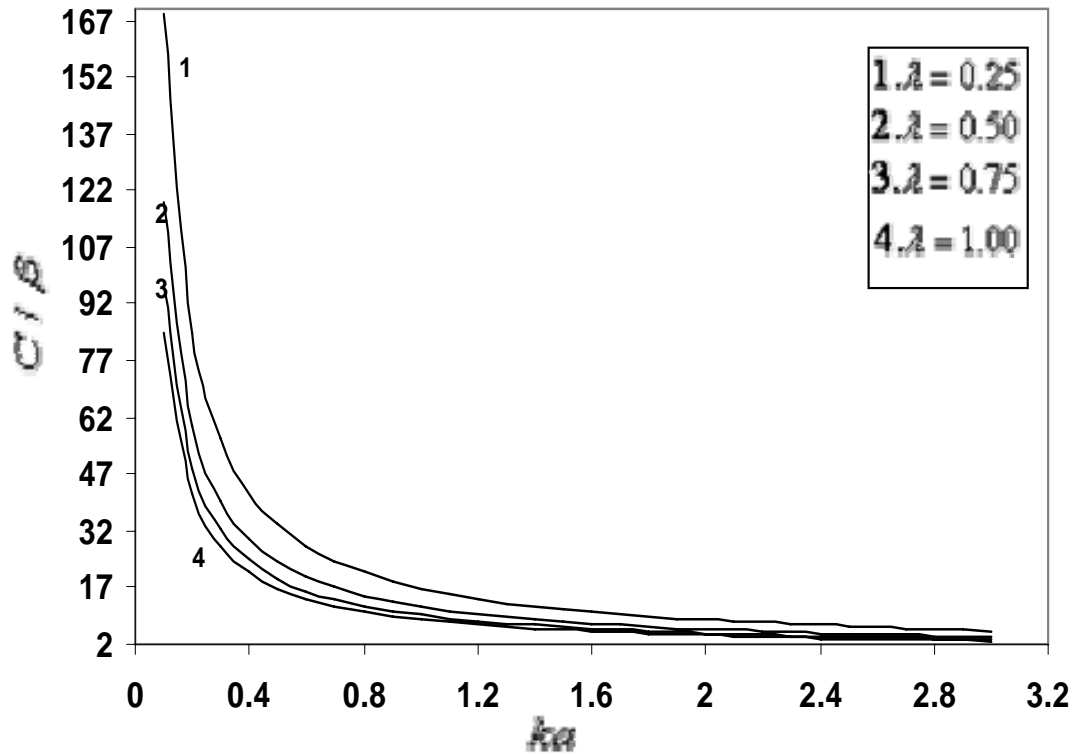

**Fig. 3 Variation of** $c/\beta$ **versus** $ka$ **at** $\xi = 8.418$

## References

[1] Bhattacharya, R. C, On the torsional wave propagation in a two-layered circular cylinder with imperfect bonding, Proc. Indian natn. Sci. Acad., 41, A, No.6 (1975),613-619.
[2] Biot, M. A., Mechanics of incremental deformation, John Wiley and Sons Inc., New York, 1965.
[3] Bromwich, T. J. L. A., On the influence of gravity on the elastic waves and in particular on the vibration of an elastic globe, Proc. London Math. Soc., 30(1898), 98- 120.
[4] Cauchy, A. L. , Exercises de mathematique. vol. 2, Bure Freres, Paris,1827.
[5] Dey, S. and Dutta, A., Torsional wave propagation in An initially stressed cylinder, Proc. Indian natn. Sci. Acad., 58, A, No.5.(1992),425-429.
[6] Ewing, W.M, Jardetzky, W. S. and Press, F., Elastic waves in layered media, McGraw-Hill, New York,1957.
[7] Kolsky, H. ,Stress wave in solids, Oxford University Press, London,1953.
[8] Lord Rayleigh ,The theory of sound, Dover, New York,1945.
[9] Love, A. E. H., The mathematical theory of elasticity. Cambridge University Press, 1927.
[10] Paul, H. S. and Sarma, K. V., Torsional wave propagation in a finite piezoelectric cylindrical shell, Proc. Indian natn. Sci. Acad., 43, A, No.2.(1977),169-181.
[11] Redwood, M. ,Mechanical Wave-guides, Pergamon . Press,1962.
[12] Southwell, R. V., On the general theory of elastic stability ,Phil.Trans.R.Soc.London, Ser. A, 213, (1913), 187-244.